\documentclass[cameraready]{Interspeech}

\title{Conversational Speech Naturalness Predictor}

\author[affiliation={1}]{Anfeng}{Xu{$^\dagger$}}
\author[affiliation={2}]{Yashesh}{Gaur}
\author[affiliation={2}]{Naoyuki}{Kanda}
\author[affiliation={2}]{Zhicheng}{Ouyang}
\author[affiliation={2}]{Katerina}{Zmolikova}
\author[affiliation={2}]{Desh}{Raj}
\author[affiliation={2}]{Simone}{Merello}
\author[affiliation={2}]{Anna}{Sun}
\author[affiliation={2}]{Ozlem}{Kalinli}

\address{
    $^1$ University of Southern California, USA. \\
    $^2$ Meta Platforms, Inc, USA.
}

\email{anfengxu@usc.edu, yagaur@meta.com}

\keywords{Speech Naturalness Estimator, MOS, Full Du-
plex, Dialogs}

\usepackage{comment}
\usepackage{multirow}
\usepackage{threeparttable}
\usepackage{booktabs}
\usepackage{graphicx}
\usepackage{tikz}

\begin{document}

\maketitle

\begingroup
\renewcommand{\thefootnote}{\fnsymbol{footnote}}
\setcounter{footnote}{0}
\footnotetext[2]{Work done while the author was an intern at Meta Platforms, Inc.}
\endgroup

\begin{abstract}
Evaluation of conversational naturalness is essential for developing human-like speech agents. However, existing speech naturalness predictors are often designed to assess utterances from a single speaker, failing to capture conversation-level naturalness qualities. In this paper, we present a framework for an automatic naturalness predictor for two-speaker, multi-turn conversations. We first show that existing naturalness estimators have low, or sometimes even negative, correlations with conversational naturalness, based on conversational recordings annotated with human ratings. We then propose a dual-channel naturalness estimator, in which we investigate multiple pre-trained encoders with data augmentation. Our proposed model achieves substantially higher correlation with human judgments compared to existing naturalness predictors for both in-domain and out-of-domain conditions.
\end{abstract}

\section{Introduction}
One of the most important aspects of a human-like speech agent is the perceived naturalness of speech, which directly affects user experience and engagement.
While manual human evaluations provide reliable judgments, they are costly and difficult to scale for iterative system development. This underscores the need for automatic predictors of conversational naturalness that can provide efficient, consistent evaluation.

Existing speech naturalness predictors have shown promising results for single-speaker, single-utterance evaluation, focusing primarily on audio quality and sentence-level fluency \cite{lo2019mosnet, mittag2021nisqa, cooper2022generalization,  baba2024t05}. However, these approaches may not extend to the assessment of conversations, where naturalness depends on 
conversational phenomena such as smooth turn-taking, the use of appropriate filler words, and the suitability of expressions based on conversation dynamics.
As a result, applying existing models directly to conversational speech can yield misaligned outcomes. This misalignment underscores the need for predictors explicitly designed to capture conversation-level naturalness.

In parallel, advances toward fully duplex conversational models, which can listen and speak simultaneously, highlight the need for robust evaluators for conversational speech generation models \cite{defossez2024moshi, zhang-etal-2025-omniflatten, hu2025efficient}. As model developments advance toward seamless agentic interactions, automatic predictors of conversational naturalness become increasingly essential for guiding model design and enabling fair performance comparisons.


To this end, we propose a simple and effective framework for automatically assessing naturalness in multi-speaker, multi-turn interactions. Our approach leverages strong pre-trained encoders to predict conversation-level naturalness with high correlation to human judgment.
Our contributions are as follows:
\begin{itemize}
    \item We conduct experiments on a large-scale synthetic conversational dataset comprising approximately 6.6k conversations with 100 hours of audio, along with an out-of-domain full-duplex conversational dataset for robustness evaluation.

    \item We show that existing predictors fail to capture conversational naturalness, often yielding negative correlations.

    \item We design a simple and effective conversational naturalness predictor framework that utilizes two-channel audio input and can leverage pre-trained speech encoders.
    \item We demonstrate that Whisper-based models exhibit the highest correlations with human judgments.

    \item We explore large-scale synthetic augmentation to further improve robustness under distribution shifts.

\end{itemize}

\section{Related Works}
\subsection{Existing Speech Naturalness Predictors}
Mean opinion score (MOS) \cite{streijl2016mean} is a widely used metric for subjective speech quality or naturalness evaluation. Existing corpora have multiple annotators rate the generated audio on a scale of 1 to 5, and the mean score is used as the human-annotated naturalness target \cite{mittag2021nisqa, maniati2022somos, huang2024voicemos}. Recent progress in deep learning has accelerated the development of automated speech quality MOS predictions \cite{lo2019mosnet, mittag2021nisqa, cooper2022generalization,  baba2024t05}. Additionally, researchers have investigated the use of audio large language models (LLMs) for the MOS prediction task \cite{wang2025enabling, zezario2025study, wu2024modelling, chen2025audio}. However, these works focus on speech quality and naturalness estimation for single-speaker utterances, and their applicability for conversational speech naturalness remains unclear. 

\subsection{Conversational Speech Generation and Evaluation}
Recent research has begun addressing speech generation from a conversational perspective, with corpora such as RyanSpeech \cite{zandie2021ryanspeech} and DailyTalk \cite{lee2023dailytalk} enabling context-aware text-to-speech (TTS). In addition, advances in spoken dialog systems have led to fully duplex (FDX) models that can listen and speak simultaneously, supporting overlaps, interruptions, and backchannels beyond strict turn-taking \cite{defossez2024moshi, zhang-etal-2025-omniflatten, hu2025efficient}. To evaluate such systems, benchmarks such as Full-Duplex-Bench v1.5 \cite{lin2025full} and Taking Turns \cite{arora2025talking} have been proposed. However, they primarily target specific phenomena such as overlap handling and turn-taking rather than holistic conversational naturalness.

\section{Dataset}

\begin{table}[t]
\footnotesize

  \caption{Datasets used in our study. The numbers under Train, Dev, and Eval indicate the number of conversation samples.}
  \label{tab:dataset}
  \centering
  \vspace{-3mm}
  \begin{tabular*}{\linewidth}{l c c c c}
    \toprule
    \textbf{Dataset} & \textbf{Rating} & \textbf{Train} & \textbf{Dev} & \textbf{Eval} \\
    \cmidrule(lr){1-1} \cmidrule(lr){2-2} \cmidrule(lr){3-3} \cmidrule(lr){4-4} \cmidrule(lr){5-5}
    ConvTTS & conversation naturalness & 4,579  & 1,000 & 1,000 \\ 
    ConvTTS & system naturalness & 3,263 & 711 & 717 \\ 
    FDX-Conv & system naturalness & 0 & 0 & 490 \\

    \bottomrule
  \end{tabular*}
\end{table}

\begin{figure}[t]
  \centering
  \centerline{\includegraphics[width=1\linewidth]{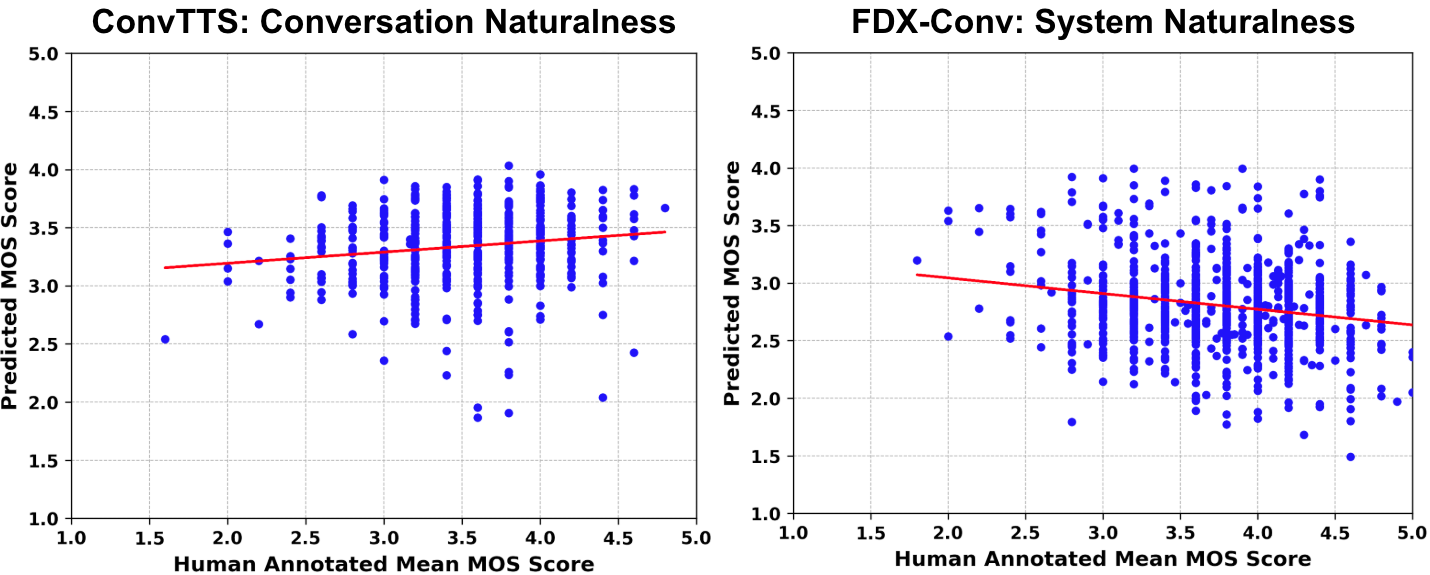}}
  \caption{Scatter plots with UTMOSv2 mean pooling methods.}
  \label{fig:utmos}
\end{figure}

\subsection{Overview}
We conduct our experiments using two internal datasets, ConvTTS and FDX-Conv (Table~\ref{tab:dataset}), both of which are two-channel recordings from the user and the conversational agent system. Each recording in the dataset is rated by at least 5 human raters for its naturalness based on 1 (not natural at all) to 5 (indistinguishable to humans) on a 1-point scale.

The naturalness of the conversation is evaluated from two perspectives, which we refer to as conversation naturalness and system naturalness. {\bf Conversation naturalness} measures the human-likeness of the conversation between the two speakers. Meanwhile, {\bf system naturalness} measures the human-likeness of the audio channel corresponding to the conversational agent system. The raters were asked to rate system naturalness by listening to both the user and system channels; therefore, system naturalness is influenced by conversational phenomena such as the naturalness of turn-taking, fillers, etc.


\subsection{Conversational TTS (ConvTTS) Dataset}
The ConvTTS dataset is a collection of dual-channel conversations generated by an internal offline TTS model.
The dataset contains 6,579 total conversation samples, with an average duration of 49.4 seconds and a standard deviation of 19.3 seconds. The generated conversations are structured in a user–system dialogue format. We randomly split the data into training, development, and test sets with 4,579, 1,000, and 1,000 samples, totaling 62.8, 13.7, and 13.8 hours, respectively. All data contain conversation naturalness ratings, while only 3263 train, 711 dev, and 717 test sets have system naturalness ratings. We use this dataset for model training and the in-domain evaluation.

\subsection{Fully Duplex Conversational (FDX-Conv) Dataset}
The FDX-Conv dataset is a collection conversations between the internal full-duplex speech generation model and internal workers.
The dataset contains 490 conversation recordings, with a mean of 41.1 seconds and a standard deviation of 19.0 seconds. Each recording is rated for system naturalness.
We use this dataset for the out-of-domain (OOD) evaluation.

\begin{table}[t]
\footnotesize

  \caption{NISQA predictions against human annotations (PCC). Utterance-level predictions are aggregated using Mean, Min, Max, and Median statistics.}
  \label{tab:nisqa}
  \centering
  \vspace{-3mm}
  \begin{tabular*}{\linewidth}{l c c c c c}
    \toprule
    \textbf{Dataset} & \textbf{Naturalness} & \textbf{Mean} & \textbf{Min} & \textbf{Max} & \textbf{Median} \\
    \cmidrule(lr){1-1} \cmidrule(lr){2-2} \cmidrule(lr){3-6} 
    ConvTTS & conversation & -0.18 & -0.13 & -0.18 & -0.18 \\ 
    ConvTTS & system & -0.25 & -0.15 & -0.22 & -0.26 \\ 
    FDX-Conv & system & 0.09 & 0.24 & -0.10 & -0.03 \\

    \bottomrule
  \end{tabular*}
\end{table}

\begin{table}[t]
\footnotesize

  \caption{UTMOSv2 predictions against human annotations (PCC). Utterance-level predictions are aggregated using Mean, Min, Max, and Median statistics.}
  \label{tab:utmos}
  \vspace{-3mm}
  \centering
  \begin{tabular*}{\linewidth}{l c c c c c}
    \toprule
    \textbf{Dataset} & \textbf{Naturalness} & \textbf{Mean} & \textbf{Min} & \textbf{Max} & \textbf{Median} \\
    \cmidrule(lr){1-1} \cmidrule(lr){2-2} \cmidrule(lr){3-6} 
    ConvTTS & conversation & -0.19 & -0.16 & -0.18 & -0.18 \\
    ConvTTS & system & -0.20 & -0.11 & -0.23 & -0.20 \\
    FDX-Conv & system & 0.15 & 0.31 & -0.02 & 0.06 \\

    \bottomrule
  \end{tabular*}
\end{table}

\section{Existing naturalness predictors}

We first evaluated whether existing naturalness predictors could assess conversational naturalness, focusing on NISQA \cite{mittag2021nisqa} and UTMOSv2 \cite{baba2024t05}. 
NISQA is a MOS predictor that consists of a CNN and a self-attention mechanism. It can predict MOS as well as perceptual dimensions such as noisiness, coloration, discontinuity, and loudness.
We use its overall MOS prediction score. UTMOSv2 combines SSL speech features and spectrogram features to predict MOS. It achieved competitive performance on the 2024 VoiceMOS challenge \cite{huang2024voicemos}. 

Unlike our goal of assessing naturalness across multi-speaker, multi-turn conversations, existing predictors are designed to assess naturalness from a single sentence spoken by a single speaker. 
Therefore, we segmented the recordings based on utterances and applied the naturalness predictors to each utterance. We then aggregated the utterance-level predictions using the mean, min, max, or median.
Conversation naturalness was computed from aggregated scores from both the user and system channels. Meanwhile, system naturalness was computed based on the aggregation from the system channel.

Figure~\ref{fig:utmos} shows the scatter plots with the UTMOSv2 mean statistic. 
Tables~\ref{tab:nisqa} and \ref{tab:utmos} show the Pearson correlation coefficient (PCC) between predicted and human-annotated MOS scores for NISQA and UTMOSv2, respectively. Both predictors exhibited negative correlations with conversation naturalness across all statistics. For system naturalness, they showed negative correlations with ConvTTS across all statistics, but positive correlations with the Mean and Min statistics for FDX-Conv. We reason that these negative correlations arose because the predictors tended to assign lower naturalness scores to more ``chit-chatty'' speech and higher scores to more formal-sounding speech, whereas human annotators judged the former as more natural within the conversational context. The positive correlation with FDX-Conv, especially for the min statistic, likely reflects cases where a severely degraded utterance lowered both predictor scores and human ratings for the entire conversation.




\section{Method}
\subsection{Modeling}

\begin{figure}[t]
  \centering
  \vspace{-2mm}
  \centerline{\includegraphics[width=0.85\linewidth]{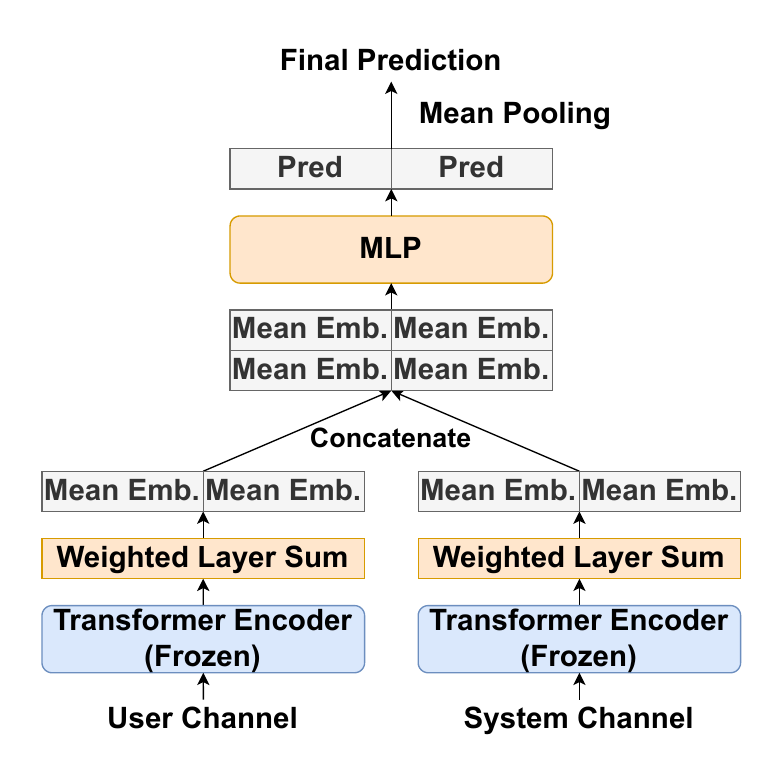}}
  \vspace{-4mm}
  \caption{Modeling architecture with dual-channel input. For the single-channle input, we only use the system-channel input.}
  \label{fig:modeling}
  \vspace{-3mm}
\end{figure}

Figure~\ref{fig:modeling} illustrates the architecture of the proposed naturalness predictors with dual-channel input modeling, which uses audio from both the user and the system. We also evaluate single-channel input modeling, which uses only system audio. Pre-trained transformer encoders are employed for feature extraction, and we compute a weighted sum of all hidden layers, followed by mean pooling over each 30-second segment. In the dual-channel setting, user and system embeddings are concatenated along the feature dimension before being passed to Multi-layer Perceptron (MLP) layers for MOS prediction. Predictions for each 30-second segment are averaged to produce the final MOS score.

For transformer encoders, we explore the following models.

\noindent\textbf{WavLM}: a self-supervised speech model that applies k-means clustering for speech quantization and is pre-trained on masked prediction and denoising tasks \cite{chen2022wavlm}. It achieves strong performance across speech recognition and understanding benchmarks. In our work, we use the WavLM-large model, pre-trained on 94k hours of audio.
 
\noindent\textbf{Audiobox-Aesthetics (AES)}: a model for automatic evaluation of audio quality across speech, music, and sound \cite{tjandra2025meta}. Trained on 562 hours of annotated data, it predicts scores for production quality, complexity, content enjoyment, and usefulness, using the WavLM-base model as its backbone.

\noindent\textbf{Whisper}: a transformer-based encoder–decoder model trained with weak supervision on 680K hours of multilingual speech data \cite{radford2023robust}. It achieved strong performance across various ASR benchmark datasets and also showed competitive results on other tasks, such as speech emotion recognition and dialect detection \cite{feng2025vox, feng2025voxlect}. We only use the encoder part for feature extraction. Specifically, we select the Whisper large-v3 model.

\subsection{Training and Evaluation Data}
We used the ConvTTS dataset for training and for the in-domain test set. Predictors were trained for both conversation naturalness and system naturalness, and the same target labels were used during testing. 
Additionally, we used the FDX-Conv dataset as an OOD test set. Specifically, we evaluated the models trained on the ConvTTS dataset for system naturalness directly on the FDX-Conv dataset. For the evaluation metric, we used PCC and Spearman's rank correlation (SRC) between predicted and ground-truth MOS scores, as well as mean squared error (MSE). 

\subsection{Data Augmentation}
We augmented the training set using the internal TTS model that had been used to develop the ConvTTS dataset. The TTS model could take in reference audio for both the system and user channels to mimic the speaker characteristics of the reference audio, similar to zero-shot TTS \cite{le2023voicebox}. For data augmentation, we used conversation samples from the training set as reference audio and used the TTS model to generate new conversation audio using text transcripts generated by Llama-3.1-405B \cite{dubey2024llama}. The generated samples were assigned the naturalness rating corresponding to the audio used for the reference signal. We generated 5,000 hours of audio with conversational naturalness ratings. We first used them explicitly for large-scale pre-training of the naturalness estimator. The model was then fine-tuned on the real ConvTTS dataset.


\subsection{Model and Training Configurations}
We used the Adam optimizer with MSE loss for all experiments. The MLP layers consist of three intermediate layers with a hidden size of 768, a dropout rate of 0.1, and GeLU activation. For training predictors based on AES and Whisper, we used a batch size of 32 and a learning rate of 0.002, while for WavLM-based predictors, we used a batch size of 16 and a learning rate of 0.001. All models were trained for 30 epochs, and the checkpoint with the lowest validation loss computed at the end of each epoch 
was selected. For pre-training with augmented data, we used a batch size of 32 and a learning rate of 0.001 for 5 epochs, evaluating every 1,000 steps and selecting the checkpoint with the highest correlation on the development set.



\section{Results and Analysis}

\begin{table}[t]
\footnotesize

  \caption{PCC, SRC, and MSE of models trained to predict \textbf{\textit{conversation naturalness}}, evaluated against human annotations.} 
  
  \label{tab:conversation-results}
  \centering
  \vspace{-3mm}
  \begin{tabular*}{0.98\linewidth}{l c c c c c}
    \toprule
    \textbf{Test} & \multirow{2}{*}{\textbf{Encoder}} & \textbf{Input} & \multirow{2}{*}{\textbf{PCC}} & \multirow{2}{*}{\textbf{SRC}} & \multirow{2}{*}{\textbf{MSE}} \\
    \textbf{Dataset} & & \textbf{Channel(s)} & & & \\
    \cmidrule(lr){1-6} 

    \multirow{7}{*}{ConvTTS} 
    
    & \multirow{2}{*}{WavLM} & single & 0.331 & 0.279 & 0.242 \\
    & & dual & 0.409 & 0.358 & 0.227 \\
    \cmidrule(lr){2-3} \cmidrule(lr){4-6}
    & \multirow{2}{*}{AES} & single & 0.384 & 0.352 & 0.228 \\
    & & dual & 0.455 & 0.430 & 0.213 \\
    \cmidrule(lr){2-3} \cmidrule(lr){4-6}
    & \multirow{2}{*}{Whisper} & single & 0.433 & 0.406 & 0.221 \\
    & & dual & \textbf{0.482} & \textbf{0.451} & \textbf{0.208} \\
    

    \bottomrule

    \vspace{-5mm}
  \end{tabular*}
\end{table}

\subsection{Results with conversation naturalness}
Table~\ref{tab:conversation-results} shows the results when the naturalness predictors were trained for the conversation naturalness. The best-performing predictor uses the Whisper encoder with dual-channel input, achieving a PCC of 0.482. The predictors using WavLM and AES encoders with dual-channel input also achieve PCCs over 0.4. The strong PCCs underscore the effectiveness of naturalness predictors targeted for the conversational context, for which the existing predictors fail to show positive results. 

\subsection{Results with system naturalness}
\begin{table}[t]
\footnotesize

  \caption{PCC, SRC, and MSE of models trained to predict \textbf{\textit{system naturalness}} evaluated against human annotations.}
  \label{tab:system-results}
  \vspace{-3mm}
  \centering
  \begin{tabular*}{1\linewidth}{l c c c c c}
    \toprule
    \textbf{Test} & \multirow{2}{*}{\textbf{Encoder}} & \textbf{Input} & \multirow{2}{*}{\textbf{PCC}} & \multirow{2}{*}{\textbf{SRC}} & \multirow{2}{*}{\textbf{MSE}} \\
    \textbf{Dataset} & & \textbf{Channel(s)} & & & \\
    \cmidrule(lr){1-6} 
    \multirow{7}{*}{ConvTTS} 
    & \multirow{2}{*}{WavLM} & single & 0.488 & 0.491 & 0.246 \\
    & & dual & 0.517 & 0.522 & 0.236 \\
    \cmidrule(lr){2-3} \cmidrule(lr){4-6} 
    & \multirow{2}{*}{AES} & single & 0.523 & 0.523 & 0.234 \\
    & & dual & 0.542 & 0.534 & 0.227 \\
    \cmidrule(lr){2-3} \cmidrule(lr){4-6} 
    & \multirow{2}{*}{Whisper} & single & 0.550 & 0.538 & 0.226 \\
    & & dual & \textbf{0.570} & \textbf{0.560} & \textbf{0.218} \\
    
    \cmidrule(lr){1-6} 
    
    & \multirow{2}{*}{WavLM} & single & 0.305 & 0.308 & \textbf{0.335} \\
    & & dual & 0.058 & 0.056 & 0.512 \\
    \cmidrule(lr){2-3} \cmidrule(lr){4-6} 
    FDX-Conv & \multirow{2}{*}{AES} & single & 0.310 & 0.297 & 0.548 \\
    (OOD) & & dual & 0.282 & 0.272 & 0.671 \\
    \cmidrule(lr){2-3} \cmidrule(lr){4-6} 
    & \multirow{2}{*}{Whisper} & single & \textbf{0.362} & \textbf{0.358} & 0.438 \\
    & & dual & 0.290 & 0.285 & 0.633 \\
    
    \bottomrule
  \end{tabular*}
\end{table}

Table~\ref{tab:system-results} shows the results when the naturalness predictors are trained for the system naturalness. For the in-domain ConvTTS test set, we observe higher PCC and SRC scores overall than for conversation naturalness. The predictor leveraging the Whisper encoder with dual-channel input modeling achieves the best results, with a PCC of 0.57 and an SRC of 0.56. On the OOD FDX-Conv test set, the Whisper-based predictor with single-channel input modeling performs best, achieving 0.362 PCC and 0.358 SRC. The results suggest that even in the OOD setting for system naturalness, our naturalness predictor outperforms the best existing baseline (UTMOSv2) of 0.31 PCC. The MSE scores are much higher and less consistent, likely due to differences in the mean MOS scores between the two datasets. 

\begin{figure}[t]
  \centering
  \centerline{\includegraphics[width=1\linewidth]{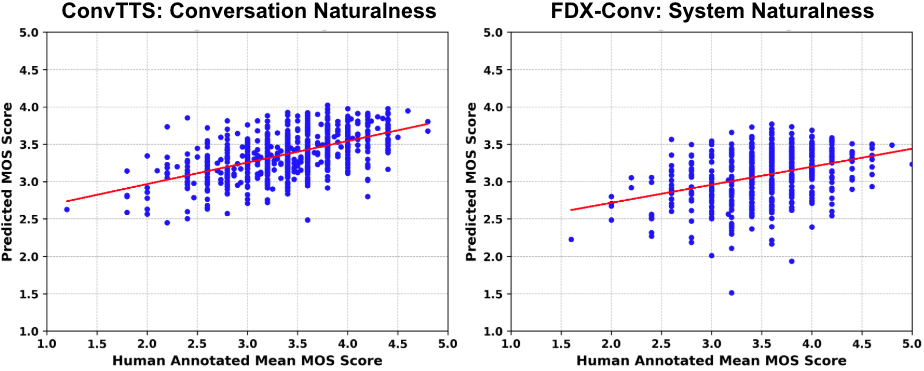}}
  \caption{Scatter plots of Whisper results. Dual-channel input for ConvTTS and Single-channel input for FDX-Conv.}
  \label{fig:scatter_whisper}
  \vspace{-2mm}
\end{figure}

\subsection{Discussion on encoder models for feature extraction}
The predictors based on the Whisper encoder consistently outperform AES or WavLM counterparts in all settings of training targets, test datasets, and input channel sizes. We reason that this is due to its large-scale pre-training of around 680k hours of audio data, enabling it to acquire more robust and diverse speech understandings, including naturalness. Between AES and WavLM, the AES-based model consistently outperforms, suggesting that the audio quality information learned through its pre-training tasks is useful for the naturalness prediction tasks. Additionally, Figure~\ref{fig:scatter_whisper} shows the scatter plots of Whisper-base modeling results, demonstrating robust positive correlations with human MOS annotations.

\subsection{Discussion on single vs dual channel input modeling}
For both conversation naturalness and system naturalness, the dual-channel input models outperform their single-channel counterparts on the ConvTTS test set. This is expected for the conversation naturalness, for which the conversational dynamics between two speakers contain crucial factors for the overall naturalness. Meanwhile, for the system naturalness, we also see that dual-channel modelings outperform single-channel modelings, indicating that information from both speakers is helpful even for single-speaker naturalness in conversational settings. 

For the FDX-Conv target test dataset, the single-channel modeling outperforms the dual-channel modeling. We reason that this is because there are substantial differences in the user speech, with ConvTTS using synthesized audio and FDX-Conv using real human recordings. This mismatch across the two datasets likely causes more noise when using the dual-channel input modeling to capture user speech.

\subsection{Data Augmentation}

\begin{table}[t]
\footnotesize

  \caption{Predictor using Whisper encoder (dual channel input modeling) with 5000 hours of augmented data pretraining for conversation naturalness. () shows the relative change against no pretraining. }
  \vspace{-3mm}
  \label{tab:augmentation}
  \centering
  \begin{tabular*}{0.83\linewidth}{l c c}
    \toprule
    \textbf{Test Dataset} & \textbf{Naturalness} & \textbf{PCC} \\
    \cmidrule(lr){1-1} \cmidrule(lr){2-2} \cmidrule(lr){3-3}
     
    \multirow{2}{*}{ConvTTS} & conversation & 0.483 (+0.21\%) \\
    & system & 0.545 (-4.39\%) \\
    \cmidrule(lr){1-1} \cmidrule(lr){2-2} \cmidrule(lr){3-3}
    FDX-Conv (OOD) & system & 0.358 (+23.45\%) \\   
    \bottomrule
  \end{tabular*}
\end{table}

Table~\ref{tab:augmentation} reports PCCs when the model is pre-trained on 5000 hours of augmented data and subsequently fine-tuned for conversation naturalness and system naturalness. While no gains are observed on the ConvTTS test set, we find 23.45\% relative improvement on the out-of-domain FDX-Conv set. We attribute this to pretraining, which provides a more generalizable initialization that enhances robustness during fine-tuning.

\subsection{Ablation on dual channel input modeling}
To test whether explicit channel separation is beneficial, we evaluated a single-channel setup where user and system audio are combined. The PCC results from Table~\ref{tab:ablation} suggest that dual-channel modeling outperforms input-channel modeling on both the in-domain ConvTTS and OOD FDX-Conv test datasets. The results show that preserving channel separation leads to stronger modeling results, suggesting that capturing user and system channels with separate encoders enhances naturalness prediction under conversational settings.

\begin{table}[t]
\footnotesize
    
  \caption{Ablation for single-channel input modeling with combined input of user and system audio, using Whisper encoder. () shows the relative change against the separated input. }
  \vspace{-3mm}
  \label{tab:ablation}
  \centering
  \begin{tabular*}{0.83\linewidth}{l c c}
    \toprule
    \textbf{Test Dataset} & \textbf{Naturalness} & \textbf{PCC} \\
    \cmidrule(lr){1-1} \cmidrule(lr){2-2} \cmidrule(lr){3-3} 
     
    \multirow{2}{*}{ConvTTS} & conversation & 0.458 (-4.98\%) \\
     & system & 0.557 (-2.30\%) \\
    \cmidrule(lr){1-1} \cmidrule(lr){2-2} \cmidrule(lr){3-3} 
    FDX-Conv (OOD) & system & 0.110 (-62.07\%) \\
                          
    \bottomrule
  \end{tabular*}
  \vspace{-3mm}
\end{table}

\section{Conclusion}
We introduced the conversational naturalness predictors for automatic evaluation of dialogue-level naturalness in multi-turn, two-speaker interactions. We showed that existing MOS predictors designed for single-utterance evaluation fail to generalize to conversational speech. To address this gap, we proposed a simple yet effective framework for predicting the naturalness of conversational speech, leveraging pre-trained speech encoders. We systematically investigated Whisper, WavLM, and AES encoders and found that Whisper-based predictors achieve the strongest correlations with human judgments. Importantly, our study revealed the critical role of dual-channel modeling in capturing interaction dynamics. 

\section{Acknowledgements}
We thank Wei-Ning Hsu and Shanyan Chen for their help in using the internal TTS model. 

\section{Generative AI Use Disclosure}
We used Generative AI tools in this study to assist with language polishing, manuscript editing, and limited code development support for data analyses and visualizations. These tools were not used to generate or interpret experimental results. All authors are aware of the extent of generative AI use, take responsibility for the content of the manuscript, and approve its submission. 

\bibliographystyle{IEEEtran}
\bibliography{mybib}

\end{document}